\documentclass{article}  
\usepackage{breckenr2005}
\usepackage{graphicx}
\usepackage{amssymb}
\frompage{000} \topage{000}                                              
\def\rightmark{Freeze-out and resonances}
\def\leftmark{L. Molnar}
 
\title{Extraction of kinetic freeze-out properties\\
and effect of resonance decays} 

\authors{
{Levente Molnar$^1$ (for the STAR Collaboration)%
}\\[2.812mm]
{\normalsize
\hspace*{-8pt}$^1$ Purdue University, Department of Physics\\ 
West Lafayette, IN, 47906\\[0.2ex] 
}}
 
\abstract{ We present STAR results from identified particle spectra measured in $\sqrt{s_{NN}}$ = 62.4 GeV Au-Au collisions. Particle production and system dynamics are compared to results at $\sqrt{s_{NN}}$ = 200 GeV. We extract kinetic and chemical freeze-out parameters using blast wave model parameterization and statistical model. We discuss the effect of resonance decays on the extracted kinetic freeze-out parameters.}

\begin{document}
 
\maketitle
\setcounter{page}{1}

\section{Introduction}\label{intro}
\begin{figure}[thb]
\resizebox{1.\textwidth}{!}{\includegraphics{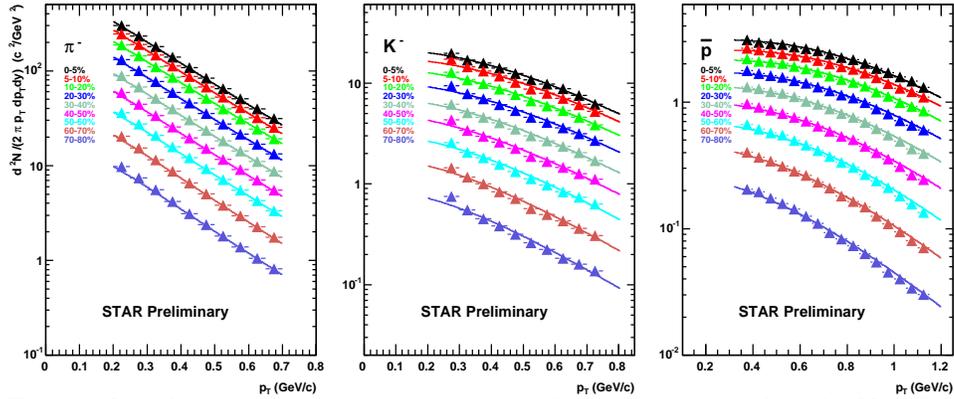}}
\vspace*{-1.cm}
\caption[]{Identified particle $p_T$ spectra of negatively charged particles in 62.4 GeV Au-Au collisions. The lines indicate the blast wave model fits without including resonance decays.}
\label{fig00}
\end{figure}

\begin{figure}[h]
\resizebox{.5\textwidth}{!}{\includegraphics{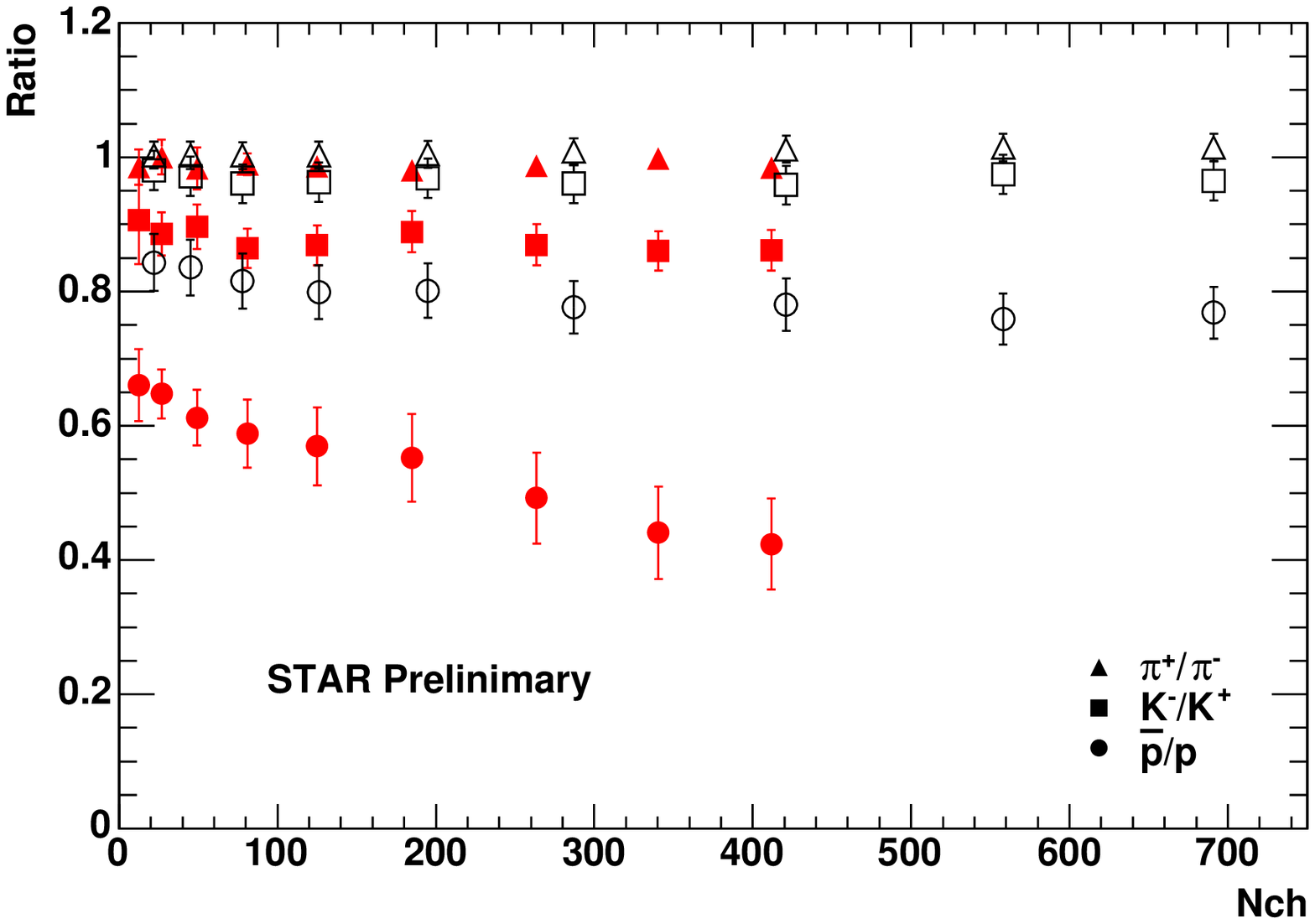}}
\resizebox{.5\textwidth}{!}{\includegraphics{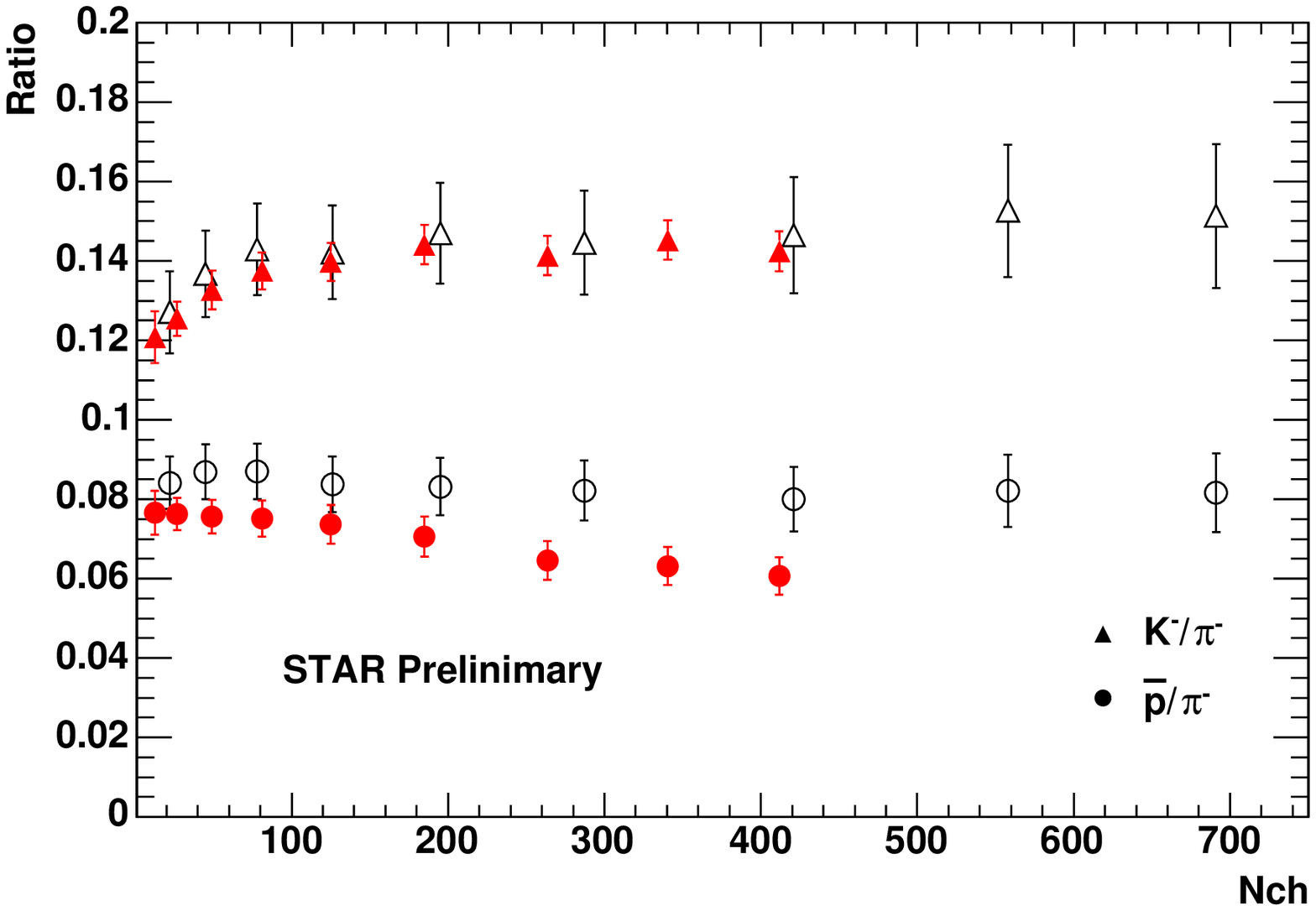}}
\vspace*{-1cm}
\caption[]{Left panel: Antiparticle-to-particle ratios at 62.4 GeV (solid markers) and 200 GeV (empty markers) as a function of centrality.
Right panel: $K^{-}$/$\pi^{-}$ and  $\overline{p}/\pi^{-}$ ratios at 62.4 GeV (solid markers) and at 200 GeV (empty markers) as a function of centrality.
 }
\label{fig01}
\end{figure}

\begin{figure}[bht]
\resizebox{.5\textwidth}{!}{\includegraphics{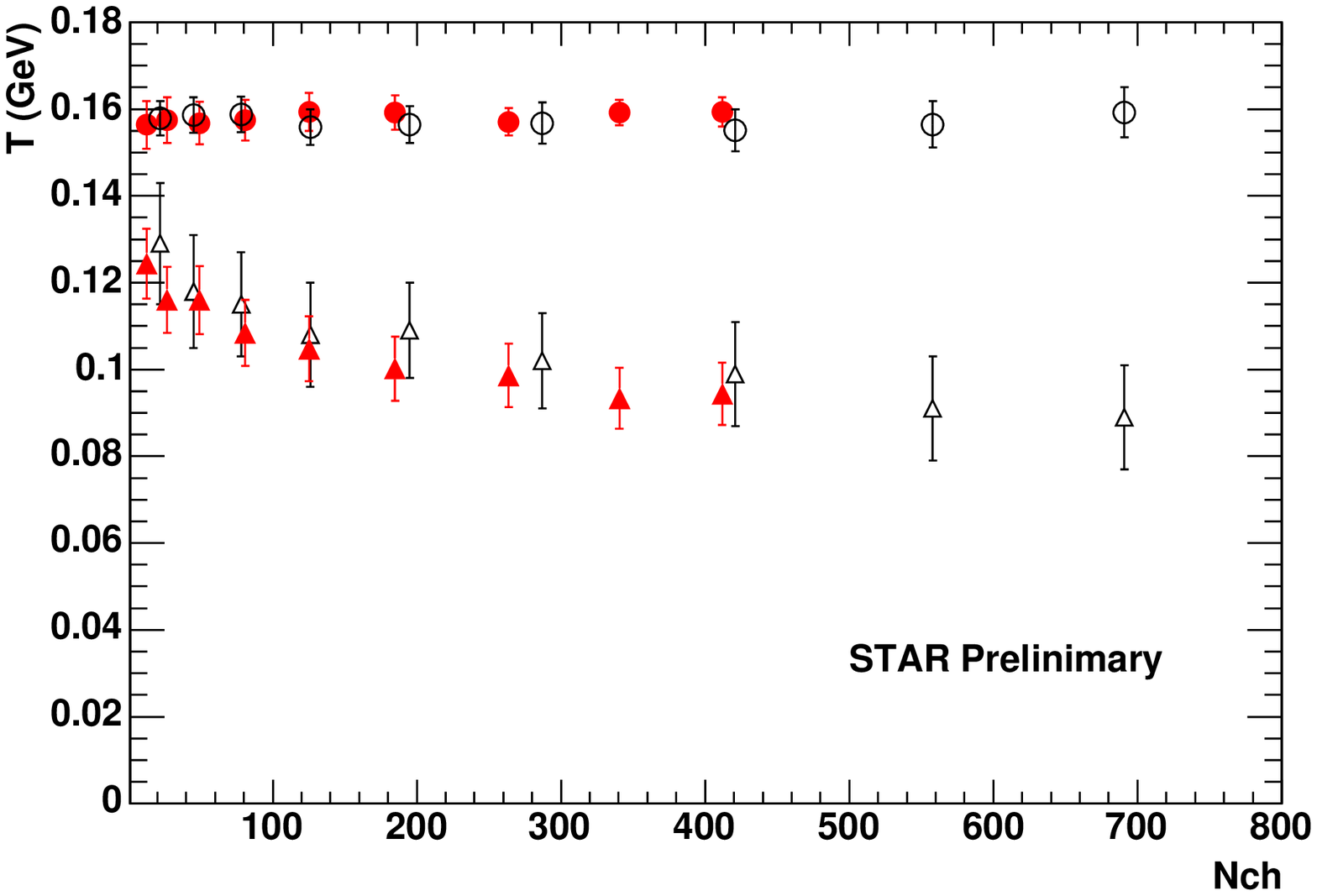}}
\resizebox{.5\textwidth}{!}{\includegraphics{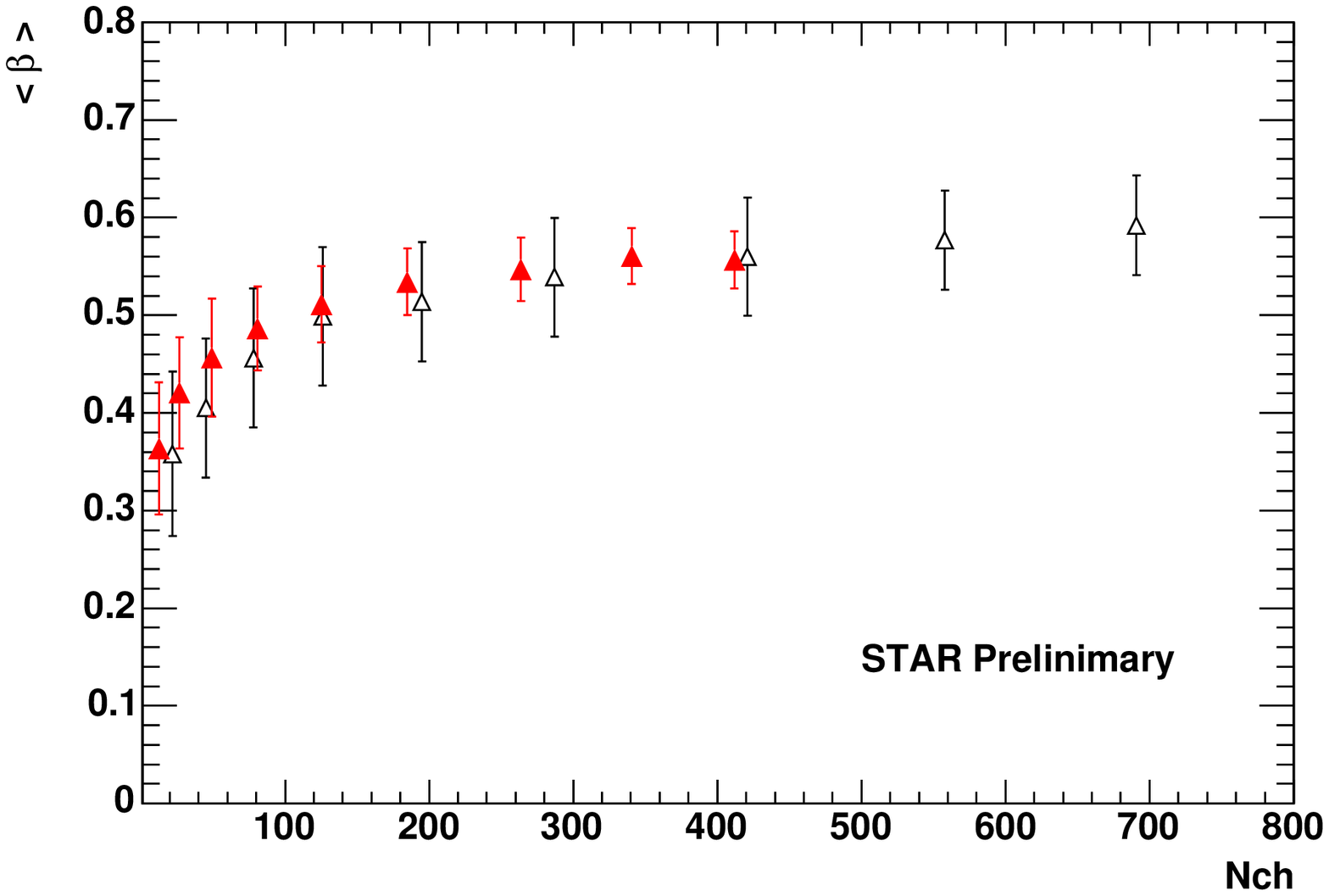}}
\vspace*{-1cm}
\caption[]{Left panel: Extracted chemical (circles) and kinetic (triangles) freeze-out temperatures as a function of centrality. The 62.4 GeV data are shown in solid markers and 200 GeV data in empty markers. Right panel: Extracted average transverse flow velocities as a function of centrality.}
\label{fig02}
\end{figure}

\begin{center}
\begin{figure}[!t]
\resizebox{0.5\textwidth}{!}{\includegraphics{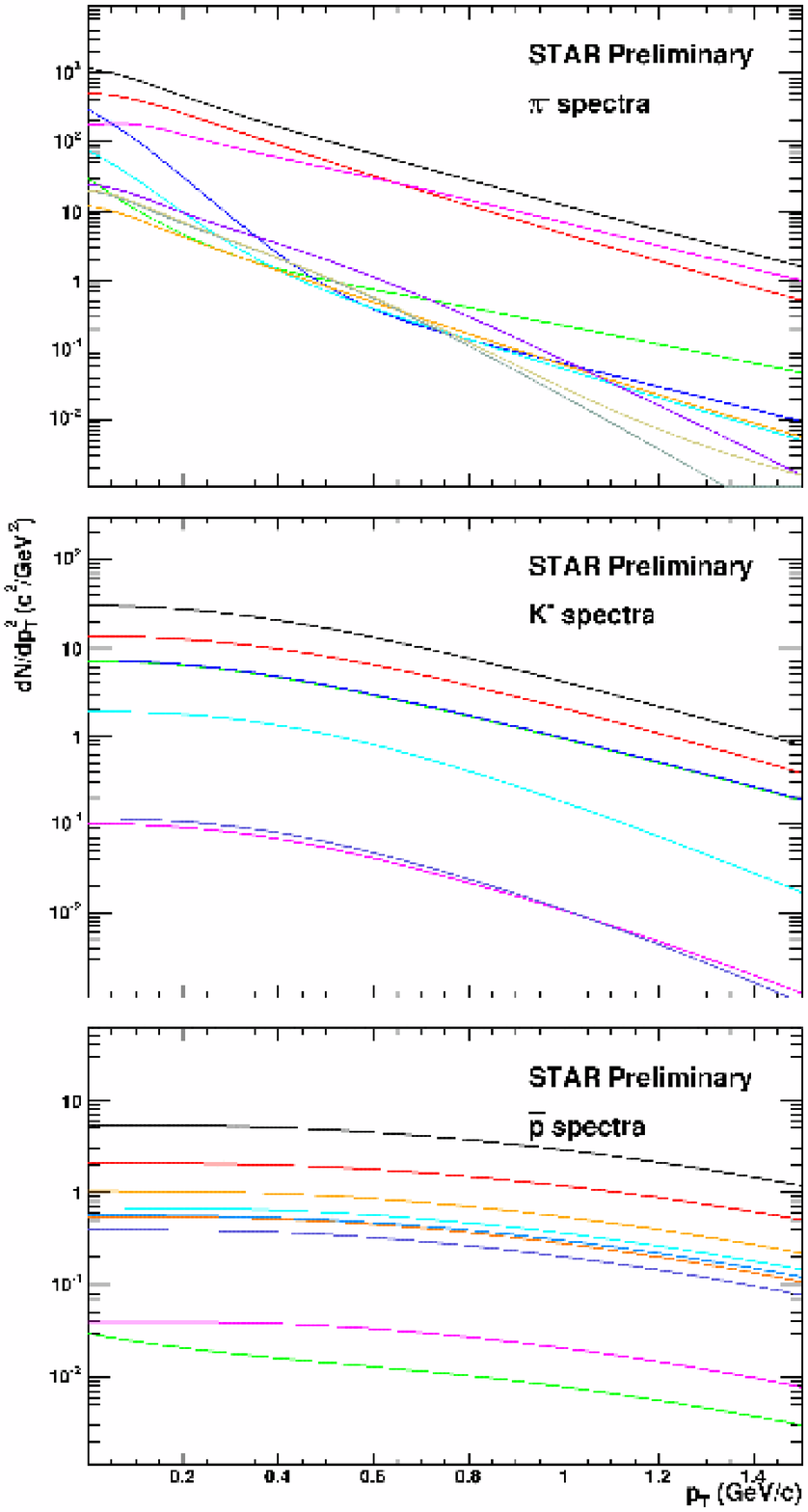}}
\resizebox{0.5\textwidth}{!}{\includegraphics{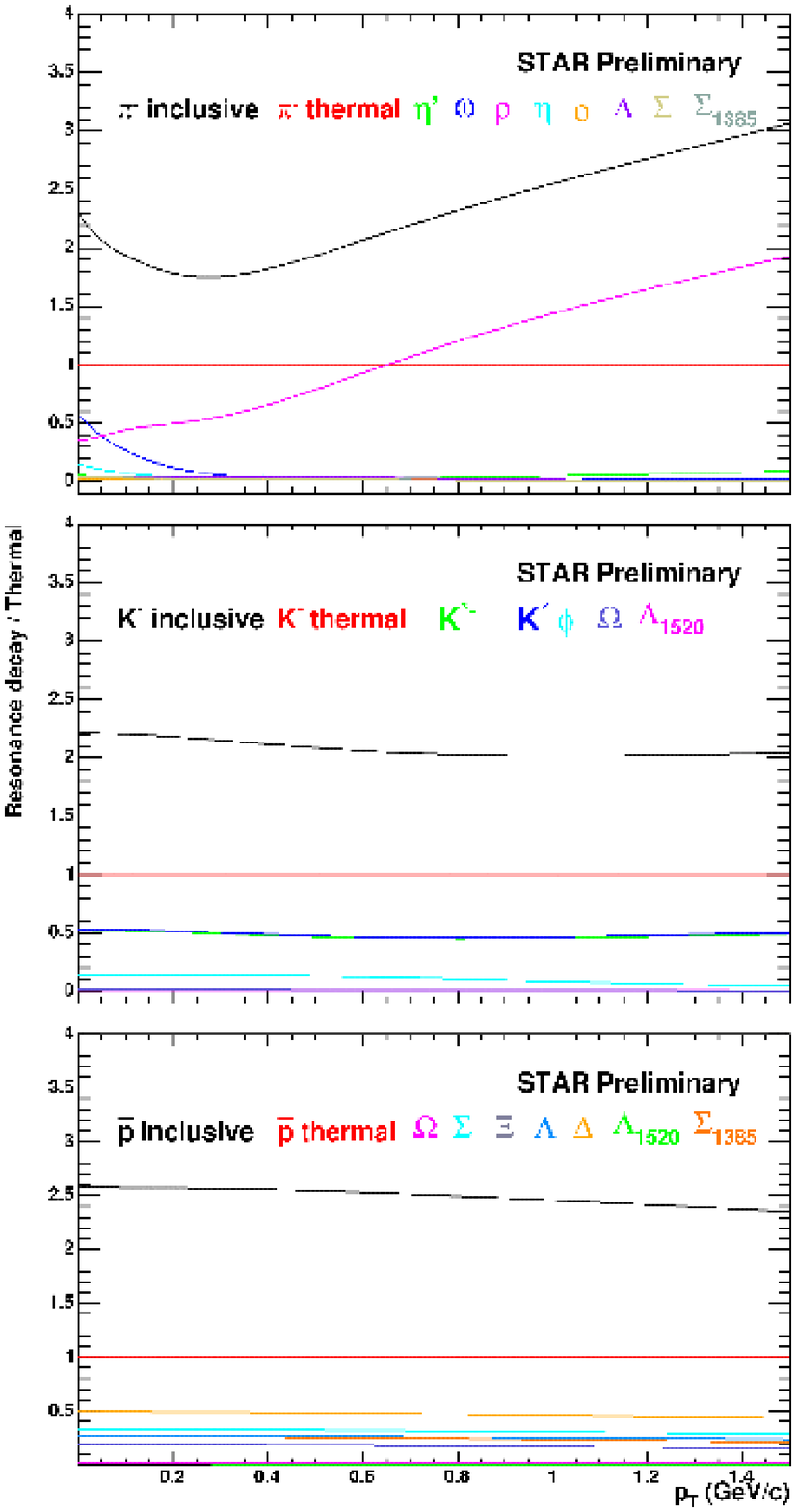}}
\vspace*{-0.1cm}
\caption[]{Left panels: Calculated contributions to $\pi^{-}$, $K^{-}$ and $\bar{p}$ from resonance decays. 
Right panels: Resonance contributions relative to the primordial spectrum.}
\label{fig03}
\end{figure}
\end{center}

\begin{figure}[!t]   
\resizebox{.5\textwidth}{!}{\includegraphics{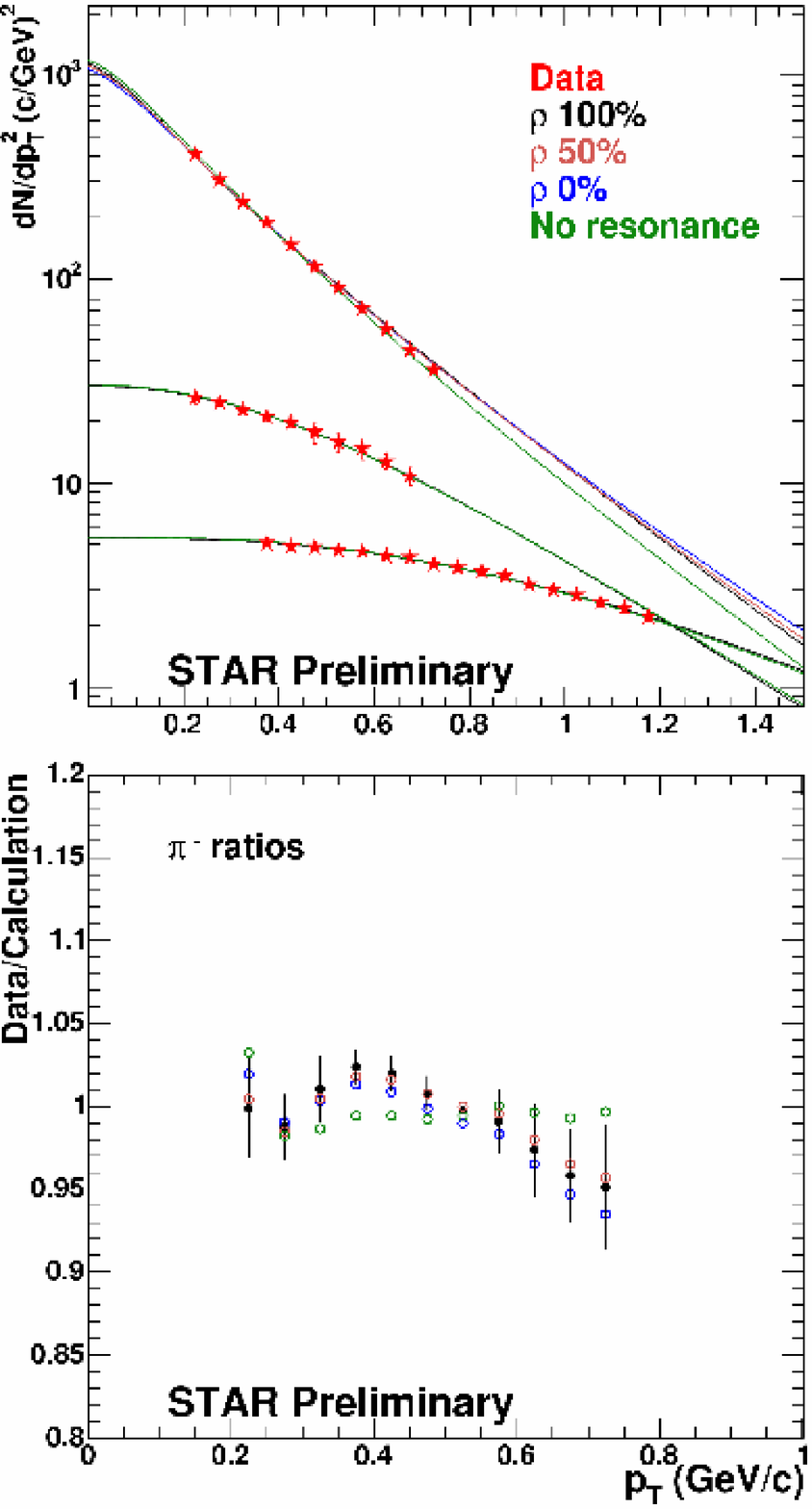}}
\resizebox{.5\textwidth}{!}{\includegraphics{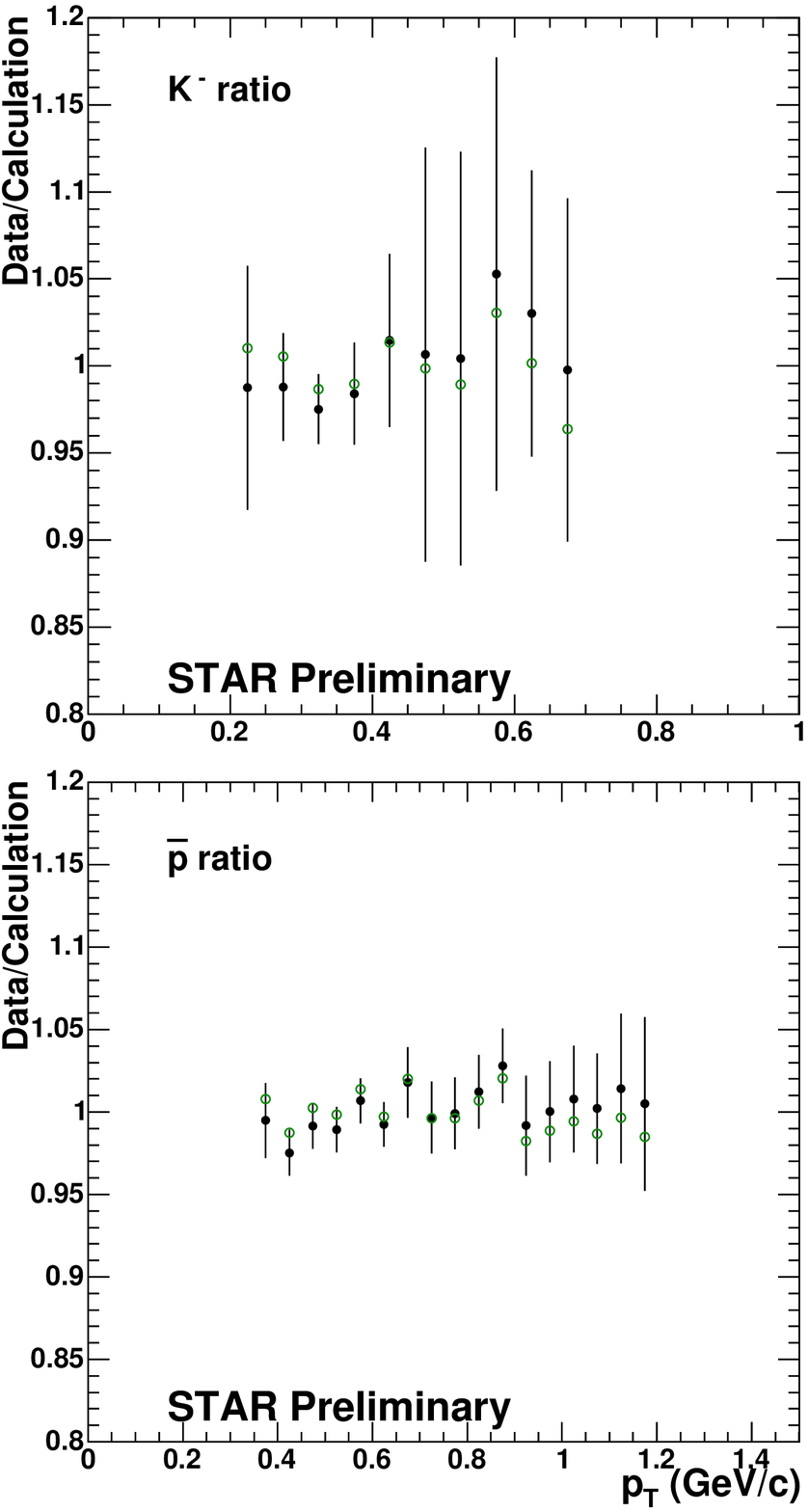}}
\vspace*{-0.1cm}
\caption[]{Top left panel: Fit of the calculated spectra to the measured ones in top 5\% central Au-Au collisions at 200 GeV \cite{Adams:2003xp}. Four calculated spectra are shown for $\pi^-$ (upper curves): including resonances with three different $\rho$ contributions and excluding resonances. Only two calculated curves are shown for $K^-$ (middle curves) and $\bar{p}$ (lower curves): including resonances with 100$\%$ $\rho$ and excluding resonances. Other panels: data / calculation ratios. Error bars are from statistical and point-to-point systematic errors on the data, and are shown for only one set of the data points.}  
\label{fig05}
\end{figure}

Results from heavy ion collisions at RHIC generated great interest in the study of nuclear matter under extreme conditions. Theory predicts the existence of a new phase of matter at sufficiently high energy density \cite{Karsch:2001vs}, which might have been reached in Au-Au collisions at RHIC. To test the hypothesis of a phase transition threshold and to map out the systematics, 
a lower energy Au-Au run has been performed at $\sqrt{s_{NN}}$ = 62.4 GeV. In this paper we investigate the bulk particle production and dynamic evolution of the colliding system and contrast the 62.4 GeV and 200 GeV results. We also address the kinetic and chemical freeze-out properties of the system and the effect of resonance decays on the extracted kinetic freeze-out parameters.

The bulk of the produced particles ($\pi^{\pm}$, $K^{\pm}$, p and $\overline{p}$) in heavy ion collision appears at low transverse momentum ($p_{T} < 2$ GeV/c). 
Bulk yield ratios of identified particles characterize the chemical freeze-out. Chemical freeze-out temperature, chemical potentials and the ad-hoc strangeness suppression factor are determined via statistical models.
Kinetic freeze-out parameters - temperature and average transverse flow velocity - are extracted from the shapes of identified particle spectra. 

\section{Data analysis}\label{techno}  
Measurements were carried out by the STAR experiment \cite{Ackermann:2002ad}.
Charged particles are detected in the Time Projection Chamber (TPC)~\cite{STARTPC}. The data sample is divided into nine centrality classes according to the measured uncorrected charged particle multiplicity in $\left|\eta\right|\leq0.5$. They represent 0-5$\%$, 5-10$\%$, 10-20$\%$, 20-30$\%$, 30-40$\%$, 40-50$\%$, 50-60$\%$, 60-70$\%$ and 70-80$\%$ of the total geometrical cross-section. 

Particles ($\pi^{\pm}$, $K^{\pm}$, $p$ and $\overline{p}$) are identified by their specific energy loss (dE/dx) in the TPC gas. Spectra are corrected for acceptance, detector and tracking inefficiencies, particle decays, and background processes. Corrections are obtained from Monte Carlo tracks embedded in real events at the raw data level with detailed description of the STAR detector geometry and realistic simulation of the TPC responses~\cite{Adler:2001bp,Adler:2001aq,Adler:2002wn}. Spectra presented here are inclusive measurements for $K^{\pm}$, $p$ and $\overline{p}$, while the pion spectra are corrected for weak decays. 
We followed the same analysis procedure as used in the measurement of identified particle spectra in 200 GeV Au-Au collisions from STAR \cite{Adams:2003xp}.

\section{Results}
Figure \ref{fig00} shows the identified particle spectra of $\pi^{\pm}$, $K^{\pm}$, $p$ and $\overline{p}$ in the $p_{T}$ range of 0.2-1.2 GeV/c, dictated by the dE/dx capability. The evolution of the particle spectra is similar to that at 200 GeV \cite{Adams:2003xp}: particle spectra harden with increasing centrality and with increasing particle mass because of collective transverse radial flow.

Due to the lower collision energy, the net baryon density at mid-rapidity is increased as compared to 200 GeV. This is shown by the $\overline{p}/p$ and $K^{-}/K^{+}$ ratios in Fig. \ref{fig01} (left panel). The $\overline{p}/p$ ratio decreases with centrality more rapidly at 62.4 GeV than at 200 GeV. Figure \ref{fig01} (right panel) shows unlike particle ratios. Qualitatively similar centrality dependence is found at the two energies. Uncorrected charged multiplicity (Nch) is chosen here to indicate centrality, however, the qualitative features of the data are the same with other centrality measures, such as the collision geometry.

Kinetic freeze-out parameters are extracted from the blast wave model without including resonances \cite{Schnedermann:1993ws}. The extracted kinetic freeze-out temperature ($T_{kin}$) exhibits a decreasing trend with increasing centrality at both energies. The average transverse flow velocity ($\langle\beta\rangle$) increases with increasing centrality.
These features are consistent with stronger expansion in more central collisions.

The blast wave parameterization is used for extrapolation of the spectra to obtain the total yields. 
Particle yield ratios are used to extract chemical freeze-out parameters using local equilibrium models \cite{Braun-Munzinger:1999qy,Xu:2001zj}. The chemical freeze-out temperature ($T_{ch}$) is independent of centrality and collision energy. The value of $T_{ch}$ is close to the QCD predicted critical temperature of phase transition \cite{Karsch:2001vs,Rischke:2003mt,chemtemp2}.

The extracted freeze-out parameters are summarized in Fig. \ref{fig02}.
The similar results between 200 GeV and 62.4 GeV indicate similar collision evolution at both energies.

\section{Effect of resonance decays}

The kinetic freeze-out parameters above were extracted without including resonance decays (i.e. treating the measured particle spectra as primordial). Pion data below 375 MeV were excluded from the fit in order to minimize effect of resonance decays. 
In this section, we address the effect of resonance decays more rigorously. For illustration, the central (0-5$\%$) 200 GeV Au-Au data is used.

Our study is based on the model by Wiedemann and Heinz \cite{Wiedemann:1996ig}, which calculates the $p_{T}$ spectra of thermal particles and decay products from resonances. There are, however, several distinctions in our study: 
(1) We assumed two distinct freeze-out temperatures while Ref.~\cite{Wiedemann:1996ig} used a single temperature. The relative proportions of particles and resonances in our study are determined by chemical freeze-out parameters that are fixed. We used $T_{ch}$= 160 MeV, $\mu_{B}$= 22 MeV, $\mu_{S}$= 1.4 MeV, and $\gamma$=0.98. In Ref.~\cite{Wiedemann:1996ig} the chemical potentials are set to zero.
(2) We included a more extensive list of resonances, namely: $\rho$, $\omega$, $\eta$, $\eta'$, $K*^0$, $K*^{\pm}$, $\phi$ and $\Lambda$, $\Delta$, $\Sigma$, $\Xi$, $\Lambda_{1520}$, $\Sigma_{1385}$, $\Omega$.
(3) Instead of Gaussian in Ref.~\cite{Wiedemann:1996ig}, we used box profile for the flow velocity: $\beta=\beta_{S}\left(r/R\right)^{n}$ where $n$ is fixed to be 0.82~\cite{Adams:2003xp}.
(4) We used a constant $dN/dy$ instead of Gaussian in Ref.~\cite{Wiedemann:1996ig}. Rapidity distributions are needed for resonances which can decay into particles at mid-rapidity where our measurements are made.

The resonance decay particle spectra are calculated, and combined with primordial ones. Spin, isospin degeneracies and decay branching ratios are properly taken into account.

The calculated particle spectra of $\pi^{-}$, $K^{-}$ and $\overline{p}$ are shown in Fig. \ref{fig03} (left panels). Since the measured pion spectra are corrected for weak decays, the calculated inclusive pion spectra do not contain weak decay pions. Resonance contributions are labeled by the initial resonance particle, e.g. a $\pi$ emerging from the $\eta' \rightarrow \eta \rightarrow \pi$ decays is labeled as $\pi_{\eta'}$. The calculated inclusive pion spectra include contributions from $\Lambda_{1520}$ which are not plotted in Fig.~\ref{fig03}.
The right panels of Fig.~\ref{fig03} show the resonance contributions to the inclusive spectra relative to the primordial one. 
The low $p_{T}$ $\pi$ enhancement is the counter play of $\rho$, $\omega$ and $\eta$; at higher $p_{T}$ the $\rho$ contribution dominates.
The inclusive kaon and antiproton spectra do not show significant changes in the spectral shapes compared to the primordial ones. The largest contributions to the inclusive kaon spectra are from $K^{*0}$ and $K^{*-}$, and the largest contributions to the inclusive $p$ and $\overline{p}$ spectra are from $\Lambda$, $\Delta$, and $\Sigma$'s. 

It is an open question what flow velocity and temperature should be assigned to the short lived resonances, such as $\rho$ and $\Delta$. These resonances are expected to be constantly regenerated during the system evolution, since their life-times are shorter than the expected system evolution time.

In other words, the processes of $\rho \rightarrow \pi\pi$ and $\pi\pi \rightarrow \rho$, for example, constantly occur along the dynamical evolution of the system. Thus, it is reasonable to expect that the final $\rho$ decay pions carry the same flow information as the primordial pions do. In other words, the regenerated $\rho$ gain negligible flow velocity during its short life span except the inherited flow from the two resonant pions.

To gain better insights, three cases are considered for $\rho$, which gives the largest contribution to the measured $\pi$ spectra.

(1) The $\rho$ decay pions have the same $p_{T}$ spectra shape as the primordial pions. (2) The $\rho$ acquires flow as given by kinetic freeze-out temperature and transverse flow velocity, and the decay pions are calculated from decay kinematics. (3) Half of the $\rho$ contribution is taken like in (1) and the other half as in (2). Case (1) has the largest flow for decay pions because the $\rho$, being heavy, acquires flow more efficiently than pions. 

Table \ref{table1} shows the fit results for the three cases. 
Also listed in Table \ref{table1} are the fit results without including resonances. The results are consistent with those in Ref.~\cite{Adams:2003xp}.

Figure \ref{fig05} shows the fits of the calculated inclusive spectra to the measured ones. Fits are performed to the six measured spectra simultaneously, but only negatively charged particles are shown. For $K^-$ and $\overline{p}$, results from the 100\% $\rho$ case fit and the fit excluding resonances are plotted, while all fits are plotted for pions.
In case of 100$\%$ $\rho$, the calculated spectrum starts to deviate from data above $p_{T}$ $\sim$ 400 MeV. In case of 0$\%$ $\rho$, below $p_{T}$ $\sim$ 400 MeV  the calculated inclusive spectrum is enhanced by $\omega$ and $\eta$, which become more important without the $\rho$.

\begin{figure}[!h]
\begin{center}
\resizebox{.8\textwidth}{!}{\includegraphics{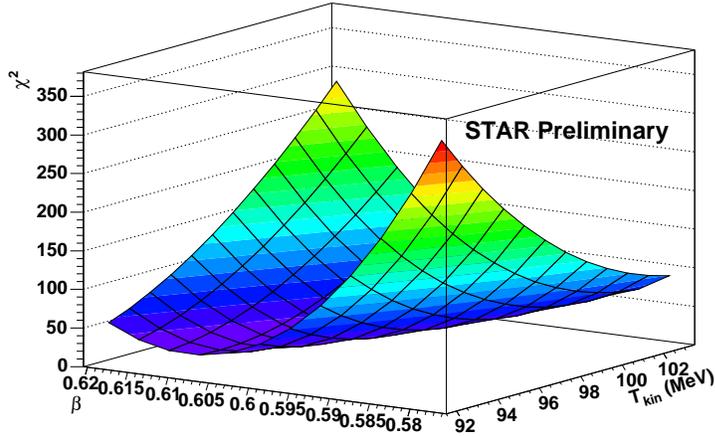}}
\end{center}
\vspace*{-0.1cm}
\caption[]{ Fitted $\chi^{2}$ as a function of $T_{kin}$ and $\langle\beta\rangle$ for the 0$\%$ $\rho$ case. Number of degrees of freedom $NDF=75$. }
\label{fig06}
\end{figure}

The model, with all the three cases of $\rho$ contributions, seems to describe the data well.
The spectra are found to be less sensitive to the kinetic freeze-out temperature than the flow velocity. 
Figure~\ref{fig06} shows, as an example, the fitted $\chi^2$ versus fit parameters $T_{kin}$ and $\langle\beta\rangle$. It is seen from the figure that $\langle\beta\rangle$ is well constrained but $T_{kin}$ is less so. Nonetheless, the fitted $T_{kin}$ values with all three cases of $\rho$ contributions seem to agree with that obtained without including resonances within systematic error of $\pm$ 10 MeV \cite{Adams:2003xp}.
In other words, resonance decays appear to have no significant effect on the extracted kinetic freeze-out parameters as shown in Table~\ref{table1}. This is primarily due to the limited $p_{T}$ ranges of our data where resonance decay products have more or less similar spectral shapes as the primordial particles do.

We also fitted the spectra data with a single, fixed kinetic freeze-out temperature $T_{kin}=T_{ch}=160$ MeV including resonances and with 100 $\%$ $\rho$ contribution. The fitted $\langle\beta\rangle$ is $0.520^{ +0.001}_{ -0.002}$ with $\chi^{2}/NDF$=19.56. A single temperature scenario is therefore ruled out by the data.

The $\chi^{2}/NDF$ is smaller than unity because we included in the fit the point-to-point systematic errors (dominate over statistical ones), which were estimated on the conservative side and might not be completely random. If we scale the $\chi^{2}/NDF$ such that the minimum is unity, then we get somewhat smaller statistical errors on the fit parameters.

\begin{table}[!b] 

\vspace*{-12pt}
\caption[]{Extracted kinetic freeze-out parameters and fit $\chi^{2}$. The flow profile n parameter is fixed to 0.82.}\label{table1}
\vspace*{-10pt}
\begin{center}
\begin{tabular}{lllll}
\hline\\[-10pt]
Set & $T_{kin}$ $(MeV)$ & $\langle\beta\rangle$ & $\chi^{2}/ndf$\\ 
\hline\\[2pt]

No resonances & $86.8^{ +0.7}_{ -0.6}$ & $0.595^{ +0.002}_{ -0.003}$  & 0.26\\[10pt]
$\rho$ 0 $\%$ & $94.6^{ +0.9}_{ -1.0}$ & $0.603^{ +0.004}_{ -0.002}$  & 0.37\\[10pt]
$\rho$ 50 $\%$& $87.4^{ +0.9}_{ -1.1}$ & $0.605^{ +0.002}_{ -0.002}$ &  0.45 \\[10pt]
$\rho$ 100 $\%$& $77.2^{ +0.8}_{ -0.9}$ & $0.604^{ +0.004}_{ -0.003}$ &  0.60\\[10pt]
\hline\\
\end{tabular}
\end{center}
\end{table}

\section{Conclusions}\label{concl}
In summary, we have reported results on identified particle spectra in 62.4 GeV Au-Au collisions and compared to those at 200 GeV. Significant drop is observed in the $\bar{p}/p$ ratio from 200 GeV to 62.4 GeV due to higher net baryon density at mid-rapidity at the lower energy. 
The extracted chemical freeze-out temperatures from statistical models show little dependence on collision centrality or energy, and coincide with the predicted phase transition temperature. The extracted kinetic freeze-out temperature drops with centrality while the radial flow velocity increases with centrality, consistent with stronger expansion in more central collisions.
The freeze-out properties show similar evolution at both energies.

We carried out a rigorous study of resonance decay effect on the extracted kinetic freeze-out parameters from the blast wave model. With different contributions from the short lived $\rho$ resonance, the extracted parameters agree with those obtained without including resonance decays within systematic uncertainties. Resonance decays appear to have no significant effect on the extracted parameters.

\section{Acknowledgment}
We acknowledge useful discussions with U. Heinz.


 
 

\vfill\eject
\end{document}